# Preparation, structural and magnetic studies on $BiFe_{1-x}Cr_xO_3$ (x=0.0, 0.05 and 0.1) multiferroic nanoparticles

Samar Layek*, Santanu Saha and H. C. Verma

Department of Physics, Indian Institute of Technology, Kanpur, 208016, India.

$BiFe_{1-x}Cr_xO_3$ (x=0.0, 0.05 and 0.1) nanoparticles are prepared by the combustion method without using any solvent. All the synthesized nanoparticles are single phase in nature, nearly spherical in shape and crystallize in distorted perovskite structure (space group *R3c*) with an average crystallite size of the order of 40 nm. The room temperature magnetization observed in $BiFeO_3$ nanoparticles is larger than that in the bulk. Saturation magnetization and coercive field increase with increasing Cr-doping. Strong superexchange interaction between $Fe^{3+}$ and $Cr^{3+}$ atoms is likely to give rise to such increase in magnetization with Cr-doping. Mössbauer data of these nanoparticles show ordered magnetic state in which Fe atoms are in 3+ oxidation states.

## I. INTRODUCTION

Multiferroic materials are compounds which show more than one (anti)ferroic properties like ferromagnetism, ferroelectricity, ferroelasticity etc [1-3]. However, there are very few materials existing in nature or prepared in the laboratory which show ferromagnetism together with ferroeletricity at room temperature or above. The reason of non-coexistence of electrical and magnetic properties in a single phase material is well discussed by Hill [4].

---

*Author to whom correspondence should be addressed

Bismuth ferrite (BiFeO₃ or commonly written as BFO) is one such example of the multiferroic material. It crystallizes in rhombohedrally distorted perovskite structure with space group *R3c*. It undergoes ferroelectric transition at 1103 K and G-type antiferromagnetic transition at the Neel temperature of 643 K [5, 6]. As both the transition temperatures are well above room temperature, it is a potential candidate for practical applications in novel magnetoelectric devices [7]. But limitations like, evolution of impurity phases during preparation due to Bi losses, poor ferroelectric properties due to oxygen vacancies and very low magnetization due to modulated spin spiral structure in addition to the G-type antiferromagnetism, restrict use of the bulk BFO in real applications [8]. To overcome these limitations, strategies based on nanotechnologies and doping on both Bi and Fe-sites of BFO have been proposed and used successfully, at least to some extent. The Fe-site doping in the BFO matrix by other transitions metals, like Cr, Mn and Ti, is reported to increase the net magnetization probably due to suppression of spin spiral structure (periodicity 62 nm) along with departure from the crystalline correlation and introduction of oxygen vacancies [9-11]. Interesting properties like cluster spin glass have been found in 5% Co-doped BFO ceramics [12]. Structural and magnetic properties of non-magnetic $Zr^{4+}$ substituted in Fe site of BFO ceramics have been described [13]. Structural, optical and transport properties of $Al^{3+}$ doped BiFeO₃ nanopowders synthesized by solution combustion method are also reported [14]. Significant increase in magnetization has been found in $BiFe_{0.95}Co_{0.05}O_3$ bulk ceramics as compared to the case of pure BFO samples prepared by rapid sintering using sol-gel derived powders [15]. Detailed magnetic and [57]Fe Mössbauer spectroscopy of Mn-doped BFO ceramics have been reported upto 10% Mn-doping. No contribution of $Fe^{2+}$ state has been found in the Mössbauer spectroscopy though there is a probability of getting $Fe^{2+}$ as a result of $Mn^{4+}$ in the Fe site, due to charge compensation [16].

Decrease in the particle size resulting in significant increase in the magnetic properties, may be due to the suppression of the spin spiral structure of periodicity 62 nm [17]. The magnetic properties in these nanosized particles greatly depend on the morphology and experimental conditions and indirectly on the preparation methods [18, 19]. Improved and interesting magnetic properties have been found in pure and doped BiFeO3 nanoparticles prepared by different methods [20-23]. Visible light photocatalytic properties have been seen in BFO nanoparticles [24]. Microwave absorption properties of $BiFeO_3$ nanoparticles have been found to be better than those of bulk BFO [25]. Gas sensing properties have also been investigated in $BiFeO_3$ nanoparticles [26].

In this paper, we report preparation of pure and Cr-doped BFO nanoparticles by a combustion method without using any solvent. The crystal structure and morphology is studied by means of x-ray diffraction (XRD) and transmission electron microscopy (TEM). The magnetic properties are investigated by vibrating sample magnetometer (VSM) and $^{57}Fe$ Mössbauer spectroscopy which has been used as a local probe to study magnetic behavior of the Fe-atoms in BFO matrix.

## II. EXPERIMENTAL DETAILS

### A. Material Synthesis

$BiFe_{1-x}Cr_xO_3$ (x=0.0, 0.05, 0.1 and 0.15) nanoparticles were prepared by a combustion method using metal nitrate and glycine as a fuel but without using any solvent. Water has been used as a solvent for preparing the solution from metal salts in many of the methods for synthesizing $BiFeO_3$ nanoparticles. But, bismuth sub-nitrate $(Bi(NO_3)_3, 6H_2O)$ is not soluble in water and turns into oxy-nitrate which may result into impurity phases after sintering along with BFO nanoparticles. A small amount of $Fe_3O_4$ phase has also been found along with $BiFeO_3$ nanoparticles by glycine nitrate combustion method using water as a solvent [18]. This small amount of $Fe_3O_4$ impurity phase resulted into higher magnetization value in these nanoparticles [18].

Bismuth nitrate $(Bi(NO_3)_3, 5H_2O)$, ferric nitrate $(Fe(NO_3)_3, 9H_2O)$, chromium nitrate $(Cr(NO_3)_2, 6H_2O)$ and glycine $(NH_2CH_2COOH)$ with purity 99.9% or higher were used without any further purification. The nitrates contain water of crystallization as one can see from their molecular formula. The idea of solution-free combustion is to mix up different nitrates at molecular scale using only this crystallization water. To do this, the nitrates were taken in proper stoichiometric ratio in a glass beaker and heated at 80 $^0C$ on a hot plate for about 10 minutes with constant stirring with cleaned glass rod. This brings out the crystallization water and the nitrates get fused to each other. Glycine was then added in appropriate proportion and mixed at that temperature for next 20 minutes. The mixture was heated at 180 $^0C$ when combustion of the mixture took place and the material turned into brown colored precursor. The precursor was grounded into fine powder using mortar and pastel. Small part of the powder was annealed using alumina crucibles in air at 600 $^0$C for 2 hours inside a programmable box furnace and then slowly cooled down to room temperature.

## B. Characterization Tools

A part of the annealed powder, for each composition, was characterized by XRD on ARL X'TRA X-ray diffractometer (Thermo Electron Corporation) using Cu- Kα radiation in order to check the crystal structure formed and calculate the average crystallite size. Crystallite size was calculated using Scherrer formula and the peak broadening of XRD peaks after correcting for the instrumental broadening. The samples for TEM measurements were prepared on a carbon-coated copper grid from a well sonicated dispersed powder in acetone. TEM micrographs were taken using a FEI Tecnai G2 electron microscope operated at 200 kV. Room temperature isothermal magnetization data were recorded using a vibrating sample magnetometer (VSM, ADE Technologies, USA) upto the maximum available magnetic field of 1.75 T. The sensitivity of the VSM instrument is quoted to be $10^{-6}$ emu. The measurements of the mass for the VSM measurements were done with an electronic balance having sensitivity of 0.1 mg. Room temperature Mössbauer spectra were recorded using a standard constant acceleration $^{57}$Fe Mössbauer spectrometer in the transmission geometry in which a source of 25 mCi $^{57}$Co in rhodium matrix was used. The data collected were analyzed using a least square fit program considering Lorentzian line-shape of the peaks and using pure iron spectrum for the calibration of velocities.

## III. RESULTS AND DISCUSSIOS

## A. Structural Analysis

Figure 1 shows the room temperature XRD patterns of pure as well as Cr-doped BiFeO3 nanoparticles. All the peaks indexed as (012), (104), (110), (006), (202), (024), (116), (022), (214), (330), (208) and (220), can be identified as coming from *R3c* structure. All the samples crystallize in the same structure as the bulk BFO sample and not much difference in the XRD patterns is seen other than decrease in the peak intensity with Cr-doping. As there is not much difference in the ionic radius of $Cr^{3+}$ (0.615Å) as compared to $Fe^{3+}$ (0.645Å) the Cr-doping does not have much effect on the crystal structure. As the ionic radius of $Cr^{3+}$ and $Fe^{3+}$ are comparable and differ from the ionic radius from of $Bi^{3+}$ it is expected that Cr atoms will substitute the Fe atoms in the BFO matrix. The average crystallite size has been calculated from the peaks indexed as (024) and (214) as other peaks are not well separated from their neighboring peak. The average crystallite size is calculated to be 44±4, 42±6 and 39±4 nm for pure, 5% and 10% Cr-doped samples respectively. The decrease in intensity with Cr-doping could be due to slight decrease in the crystallite size. It is worthy to note that the particle size is well below the period of spin spiral structure of bulk BFO and expected to show higher magnetization.

Rietveld refinements were done on all the three annealed samples based on the XRD data to further analyze the structural properties and to determine the crystal parameters using MAUD software [27]. XRD diffraction data in 2θ range $20^0$-$70^0$ has been used for the refinement. Bi atoms at x=0, y=0 and z=0, Fe atoms at x=0, y=0 and z=0.2194 and oxygen atoms at x=0.4346, y=0.0119 and z=-0.0468 crystallographic positions in the rhombohedral structure (space group *R3c*) were taken as the starting parameters for the refinement. The lattice parameters a=b=5.571 Å and c=13.858Å (taken from single crystal data from the reference [28]) were taken as the initial values and then refined for the intensity matching. For the Cr-doped samples, refinements are done assuming Cr atoms to occupy the Fe sites with initial occupancy 0.05 and 0.1 for 5 and 10% doped samples respectively. The XRD profiles of all these samples after Rietveld refinement are shown in figure 2 where blue plus signs are the experimental points and solid black lines represent calculated values. Bragg positions of the $BiFeO_3$ crystallizing in space group-*R3c* are shown by the small vertical lines. The final Cr occupancy is found to be 0.0486 and 0.0984 for 5 and 10% doped samples respectively for the best refinement of the XRD data. This shows that all the Cr is going into the expected $Fe^{3+}$ sites and no extra phase is being formed. The crystal parameters extracted from the refinement are given in Table 1. Both crystal parameters a and c decrease with increasing Cr-doping which may be due the smaller ionic radius of $Cr^{3+}$ (0.615Å) as compared to $Fe^{3+}$ (0.645Å). The data show that the decrease in the value of a is much more than that in the value of c.

Figure 3 shows TEM micrograph for 10% Cr-doped $BiFeO_3$ nanoparticles. These particles are nearly spherical in shape and agglomerate at some places. The sample for TEM grid was sonicated for quite some time. Similar procedure for many other samples prepared in our laboratory gives well dispersed particle images in TEM. The agglomeration of the particles in the present case may be arising due to strong interparticle interaction. Though a precise determination of particle size distribution is difficult because of agglomeration, the range seems to be around 35-60 nm. This agrees well with the XRD results showing the average crystallite size to be around 45 nm.

**B. Magnetic Measurements**

The magnetization curves as a function of applied magnetic field (hysteresis loop) of three samples are presented in figure 4. These hysteresis loops were measured at room temperature using vibrating sample magnetometer. The maximum field available with this machine is 1.75 tesla. Magnetization increases rapidly at low fields upto about 0.3 tesla and then increases slowly with further increase in magnetic field. The magnetization does not saturate upto the highest applied magnetic field. Typical ferromagnetic nature can be seen from the non-zero value of coercive field (shown in the inset of figure 4) for all these samples.

The value of the magnetization at higher applied magnetic field is about 0.22 emu/g in the pure BFO nanoparticles which is about double the value observed in bulk BFO prepared by conventional solid state reaction method. The increase in magnetization in these 42 nm pure nanoparticles is expected due to suppression of spin spiral structure of periodicity 62 nm. The suppression of spin spiral structure has been reported in BFO nanoparticles prepared by a chemical method [29]. The value of magnetization of the pure BFO ceramics, prepared by glycine-nitrate combustion method using water as solvent, is also higher ($\approx$ 1 emu/g at 2T) [9]. The nanoparticles prepared by the same method show a magnetization of about 4.8 emu/g at room temperature, which is much higher than our value [18]. But, the probable reason of this higher magnetization is reported to be the presence of $Fe_3O_4$ impurity phase which is highly magnetic. The nanoparticles in the present study are prepared without using water as solvent and no iron oxide impurity is found. The absence of such phases has also been confirmed by slow scan XRD (not shown here) and the Mössbauer spectroscopy discussed later.

From figure 5, we can see clearly that the magnetization increases with increasing Cr-doping upto the highest 10% doping we have studied. It is worth noting that not only magnetization but also the coercive field increases with Cr-doping. The magnetization at highest applied magnetic field (1.75 T), remanent magnetization ($M_r$) and coercive field ($H_c$) as a function of Cr-doping concentration are given in Table 1. All these three quantities increase with increasing Cr content. There can be several reasons for increase in the magnetization. Change in Fe-O-Fe bond angle due to more distorted crystal structure than BFO can be one such reason. The asymmetry in contraction of crystal parameter (a and c) values, as seen from Table-1, can result in such change in the bond angle. Suppression of spin spiral structure due to $Cr^{3+}$ doping could be another reason. Such breaking of spiral spin structure has been found in Mn-doped BFO ceramics [30]. Strong $180^0$ superexchange interaction between Cr and Fe atoms can also lead to higher magnetization [31]. The magneto-crystalline anisotropy constant (K) can be written from Stoner Wohlfarth theory [32, 33] as K= $H_c M_s \mu_0/2$ where $H_c$ indicates coercive field, $M_s$ indicates saturation magnetization and $\mu_0$ is the free space permeability. The increase in the coercive field and net magnetization indicates increase in the magneto-crystalline anisotropy with increasing Cr-doping in the systems.

### C. Mössbauer Spectroscopic Study

[57]Fe Mössbauer spectroscopy is one of the most efficient tools to investigate the local magnetic behavior and oxidation state of the iron atoms in a matrix. The Mössbauer spectra for undoped, 5% and 10% Cr-doped samples are shown in figures 5(a) to 5(c). The black dots represent the experimentally recorded data points whereas the blue lines are the least square fit of the spectrum. The parameters extracted from the best fit data are given in Table-2. There is an error of about ±0.02 mm/sec in isomer shift (IS), quadrupole splitting (QS) and linewidth whereas the errors associated with hyperfine field and area ratio is about ±0.1 T and ±2% respectively.

The spectra can be best fitted with two sextets (shown in the figure by red and green colors) of isomer shift ranging from 0.28 to 0.37 mm/s, nearly zero QS and magnetic hyperfine field of 49.3 to 49.6 tesla. We first tried to fit the data with one sextet which resulted into higher value of linewidth 0.45 mm/s (calibration spectrum of bcc iron gives linewidth of 0.26mm/s). It can be seen in figure 5 that the two-sextet fit matches reasonably well with experimental data points. Mössbauer spectroscopy is very sensitive to the iron containing phases. No iron containing phases other than $R3c$ perovskite phase is found. The results clearly indicate that the increase in the magnetization is an intrinsic property of the sample and is not resulting from any impurity phases (like $Fe_2O_3$, $Fe_3O_4$, metallic Fe, $CrFe_2O_4$ and $Bi_2Fe_4O_9$ etc.). The isomer shift which is a measure of the oxidation state of Fe atoms is found to be 0.28 to 0.37 mm/s. These values indicate that Fe-atoms are in $Fe^{3+}$ state in all the samples. There is no indication of $Fe^{2+}$ or $Fe^{4+}$ due to introduction of Cr-ions in the BFO matrix. There is practically no change in the value of isomer shifts (IS), quadrupole splitting (QS) and especially in the magnetic hyperfine field value with Cr-doping though the magnetization value increases significantly with doping. The area ratio of the sextet with higher isomer shift to that with lower isomer shift increases slightly with Cr-doping concentration. Two sextets in the Mössbauer spectra have been found in $BiFeO_3$ nanoparticles [29], Ca, and Pb-doped $BiFeO_3$ and $Bi_{0.8}La_{0.2}Fe_{1-x}Co_xO_3$ bulk samples [34-36]. On the other hand, single magnetic sextet corresponding to $Fe^{3+}$ state of the iron atom have been reported in the $BiFe_{1-x}Mn_xO_3$ (x= 0.00, 0.05 and 0.10) ceramics where magnetic hyperfine field decreases significantly with increasing Mn doping [16].

## IV. CONCLUSIONS

Pure and Cr-doped $BiFeO_3$ nanoparticles upto 10% doping have been successfully prepared by a simple combustion method. XRD results show that particles are single phase in nature and crystallize in the distorted perovskite structure with average particle size around 40 nm. Ferromagnetism with noticeable coercive field has been found in all samples. The magnetization and coercive field values increase with increasing Cr-doping. No iron containing phase other than $R3c$ perovskite phase has been detected by $^{57}Fe$ Mössbauer spectroscopy. The spectra can be fitted with two sextets corresponding to magnetic state of $Fe^{3+}$.

Table 1: The value of magnetization at highest applied magnetic field (1.75 T) ($M_S$), remanent magnetization ($M_r$), coercive field ($H_C$) and squareness (S) of $BiFe_{1-x}Cr_xO_3$ (x= 0.00, 0.05 and 0.10) nanoparticles measured at 300K using VSM magnetometer.

| Sample | Particle size (nm) | Crystal parameter a (Å) | Crystal parameter c (Å) | $M_s$ (emu/g) | $M_r$ (emu/g) | $H_c$ (Oe) | Squareness (S) |
|--------|------|------|------|------|------|------|------|
| x= 0.00 | 42 | 5.5854 | 13.8804 | 0.22 | 0.005 | 56.7 | 0.03 |
| x= 0.05 | 40 | 5.5667 | 13.8787 | 0.56 | 0.038 | 83.6 | 0.09 |
| x= 0.10 | 39 | 5.5397 | 13.8761 | 0.93 | 0.133 | 109.9 | 0.14 |

Table 2: Mössbauer parameters fitted to $BiFe_{1-x}Cr_xO_3$ (x= 0.00, 0.05 and 0.10) spectra.

| Sample | Sextet | IS (mm/s) | QS (mm/s) | LWD(mm/s) | HMF(T) | AREA (%) |
|--------|--------|-----------|-----------|-----------|--------|----------|
| x =0.00 | 1 | 0.37 | 0.03 | 0.36 | 49.6 | 52 |
|         | 2 | 0.30 | 0.02 | 0.37 | 49.4 | 48 |
| x =0.05 | 1 | 0.36 | 0.02 | 0.38 | 49.5 | 56 |
|         | 2 | 0.29 | 0.03 | 0.38 | 49.3 | 44 |
| x =0.10 | 1 | 0.34 | 0.02 | 0.39 | 49.5 | 57 |
|         | 2 | 0.28 | 0.03 | 0.40 | 49.3 | 43 |

**Figure Caption:**

Fig.1: (Color online) Room temperature powder x-ray diffraction pattern of $BiFe_{1-x}Cr_xO_3$ nanoparticles for (a) x=0.00, (b) x=0.05 and (c) x=0.10.

Fig. 2: (Color online) Rietveld refinement profile of the powder XRD pattern of (a) undoped, (b) 5% and (c) 10% Cr- doped $BiFeO_3$ nanoparticles at room temperature. Blue plus signs are the experimental points, solid black lines are from the calculated values and vertical bars are Bragg positions of the $BiFeO_3$. Difference between experimental data and theoretical fit is shown in the bottom part.

Fig.3. TEM micrograph of 10% Cr-doped $BiFeO_3$ nanoparticles synthesized by combustion method.

Fig.4: (Color online) Room temperature M-H hysteresis curve of $BiFe_{1-x}Cr_xO_3$ (x= 0.00, 0.05 and 0.10) nanoparticles. The experimental error is around $10^{-4}$ emu/g.

Fig.5: (Color online) Mössbauer spectra of $BiFe_{1-x}Cr_xO_3$ nanoparticles for (a) x=0.00, (b) x=0.05 and (c) x=0.10.

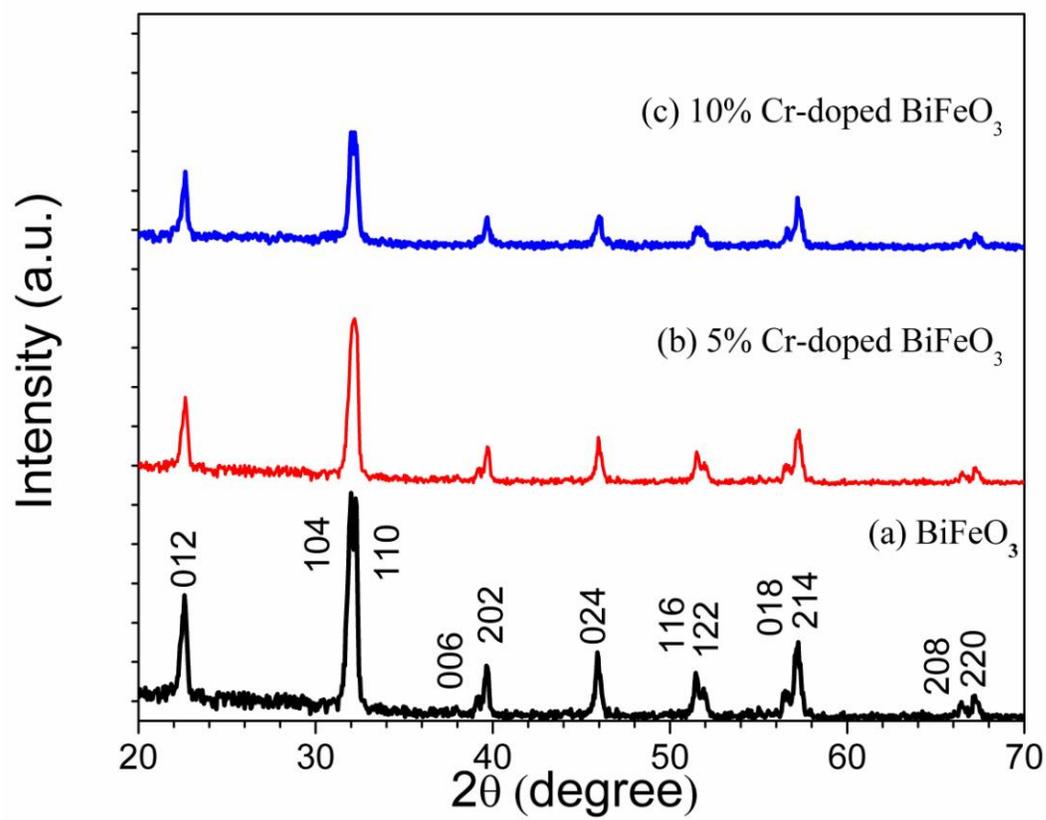

**Figure 1**

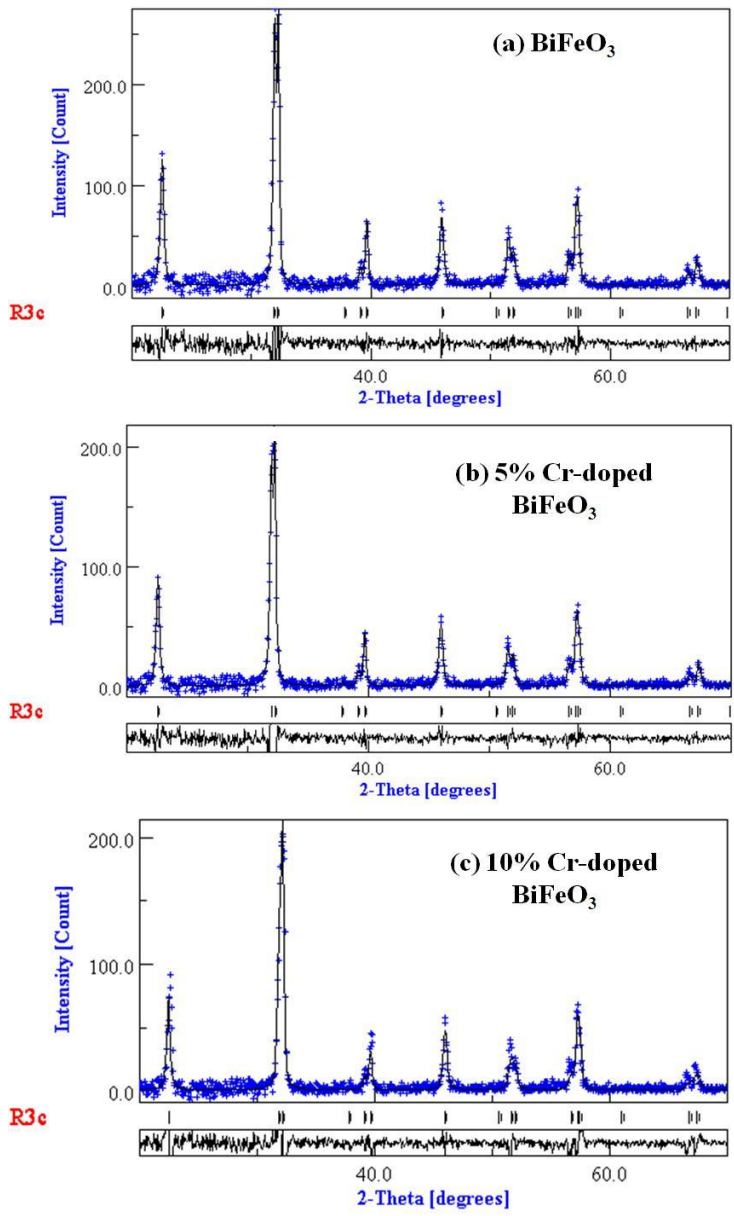

**Figure 2**

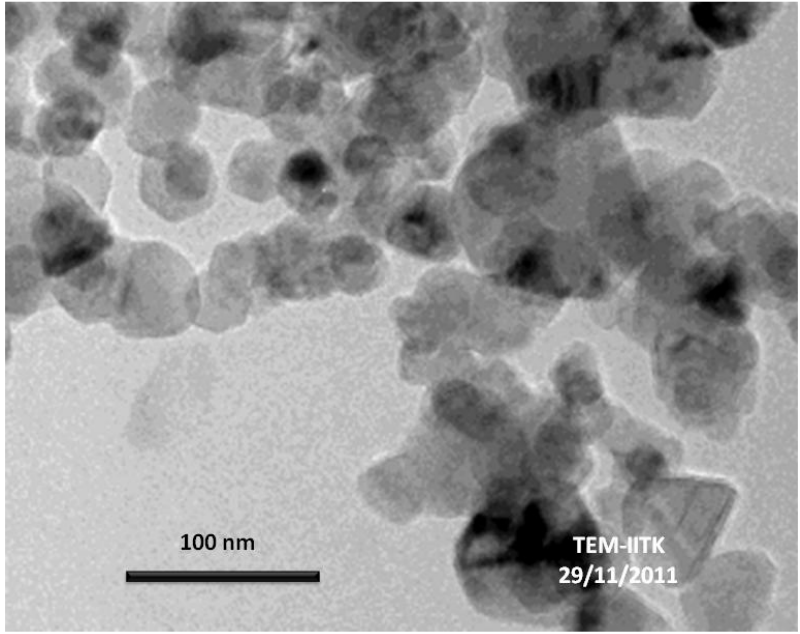

**Figure 3**

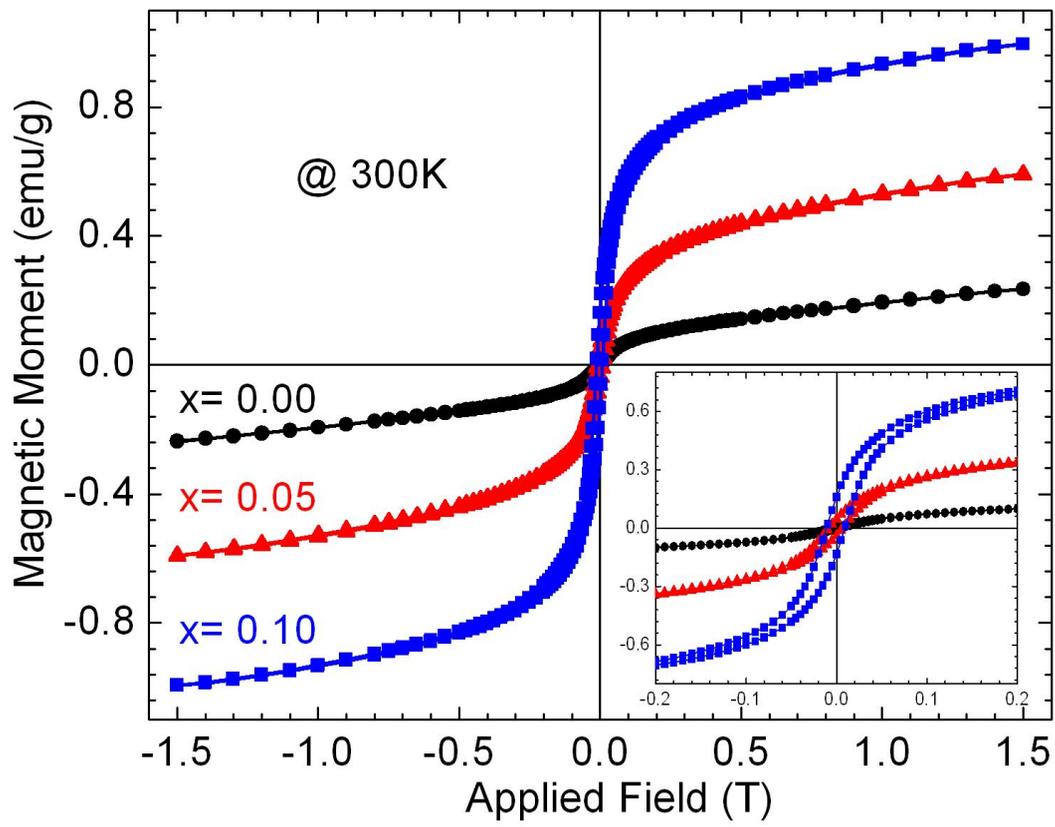

**Figure 4**

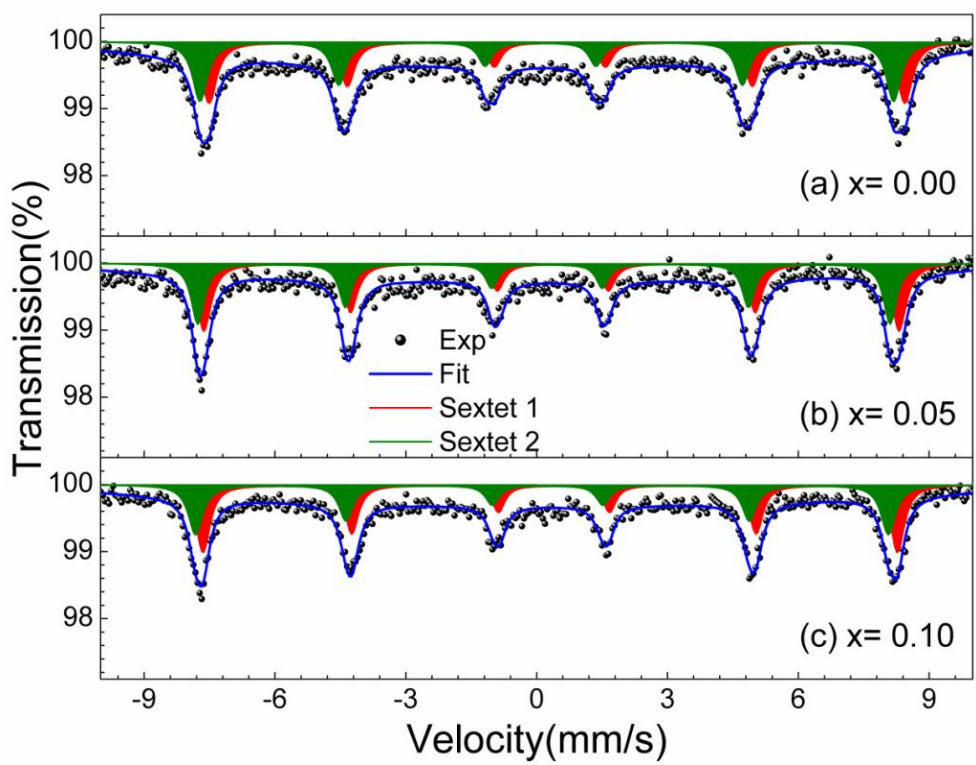

Figure 5